\documentclass[showpacs, amsmath, amssymb, twocolumn, superscriptaddress, showpacs, pra]{revtex4-1}
\usepackage{color}
\usepackage{amsfonts}
\usepackage{amsbsy}
\usepackage{latexsym}
\usepackage{mathrsfs}
\usepackage{bm}
\usepackage[dvipdfm]{graphicx}

\newcommand{\be}{\begin{eqnarray}}
\newcommand{\ee}{\end{eqnarray}}
\def\({\left(}
\def\){\right)}
\def\[{\left[}
\def\]{\right]}

\def\R{\mathbb{ R}}
\def\Z{\mathbb{ Z}}

\newcommand{\ve}[1]{ \boldsymbol #1}

\newcommand{\sla}[1]{\rlap{\kern .15em /}#1}

\newcommand{\ot}{\otimes}

\def\njp#1#2#3{{ New\ J.\ Phys.}\ {\bf {#1}}, #3 ({#2})}

\def\prA#1#2#3{{ Phys.\ Rev.\ A}\ {\bf {#1}}, #3 ({#2})}

\def\prl#1#2#3{{ Phys.\ Rev.\ Lett.}\ {\bf #1}, #3 ({#2})}

\begin{document}

% Title Page
\title{Minimal and Robust Composite Two-Qubit Gates with Ising-Type Interaction
%of Transverse Field Ising Interactions
}

\author{Tsubasa Ichikawa}
\altaffiliation[Present address: ]{Department of Physics, Gakushuin University, Tokyo 171-8588, Japan}
\affiliation{Research Center for Quantum Computing, Interdisciplinary
Graduate School of Science and Engineering, Kinki University, 3-4-1
Kowakae, Higashi-Osaka, Osaka 577-8502, Japan}
\author{Utkan G\"ung\"ord\"u}
\affiliation{Research Center for Quantum Computing, Interdisciplinary
Graduate School of Science and Engineering, Kinki University, 3-4-1
Kowakae, Higashi-Osaka, Osaka 577-8502, Japan}
\author{Masamitsu Bando}
\affiliation{Research Center for Quantum Computing, Interdisciplinary
Graduate School of Science and Engineering, Kinki University, 3-4-1
Kowakae, Higashi-Osaka, Osaka 577-8502, Japan}
\author{Yasushi Kondo}
\affiliation{Research Center for Quantum Computing, Interdisciplinary
Graduate School of Science and Engineering, Kinki University, 3-4-1
Kowakae, Higashi-Osaka, Osaka 577-8502, Japan}
\affiliation{Department of Physics, Kinki University, 3-4-1 Kowakae,
Higashi-Osaka, Osaka 577-8502, Japan}
\author{Mikio Nakahara}
\affiliation{Research Center for Quantum Computing, Interdisciplinary
Graduate School of Science and Engineering, Kinki University, 3-4-1
Kowakae, Higashi-Osaka, Osaka 577-8502, Japan}
\affiliation{Department of Physics, Kinki University, 3-4-1 Kowakae,
Higashi-Osaka, Osaka 577-8502, Japan}

%\date{\today}

\begin{abstract}
We construct a minimal robust controlled-NOT gate with an Ising-type interaction
by which elementary two-qubit gates are implemented. It is robust against inaccuracy 
of the coupling strength and the obtained quantum circuits are constructed with 
the minimal number ($N=3$) of elementary two-qubit gates and several one-qubit 
gates. It is noteworthy that all the robust circuits can be mapped to one-qubit 
circuits robust against a pulse length error. We also prove that a minimal robust 
SWAP gate cannot be constructed with $N=3$,  but requires $N=6$ elementary two-qubit 
gates. \end{abstract}
 

 \maketitle

%%%%%%%%%%%%%%%%%%%%%%%%%%%%%%
\section{Introduction}
%%%%%%%%%%%%%%%%%%%%%%%%%%%%%%

Gates robust against noise
are essential for reliable computation
irrespective of its nature being classical or quantum \cite{Feynman96, NC00, Gaitan07, NO08}.
In quantum computation, one of the promising ways 
to achieve this goal is to employ composite gates \cite{Jones11, Jones09}: 
Realizing a target gate as a sequence of elementary gates
so that systematic errors inherent in elementary gates cancel with each other.

It has been becoming clear that one must pay attention to both the noise and systematic errors,
so as to materialize a working quantum computer. Indeed, in a layered architecture for quantum 
computing, physical qubits undergo the composite gate operations and dynamical
decoupling as preliminary processing to quantum error correction, which compensates for
bit flips and phase flips due to the stochastic noise \cite{CodyJones12}.

Composite gates (pulses) have been developed in
nuclear magnetic resonance (NMR),
which is also a good test bed for quantum computation. 
Operations in NMR are implemented with the following elementary
unitary gates:
\be
R(\theta, \phi)&=&\exp[-i\theta(\cos\phi\sigma_x+\sin\phi\sigma_y)/2],\nonumber\\
S(\Theta)&=&\exp(-i\Theta\sigma_z\ot\sigma_z/4),
\label{R}
\ee
where $\sigma_i\ (i=x, y, z)$ are the Pauli matrices.
We will consider quantum circuits made of these gates (\ref{R}).
The operations
$R(\theta, \phi)$'s are realized with so-called rf-pulses, 
while $S(\Theta)$'s are implemented by free evolutions of a system 
under the Hamiltonian
$J\sum_i\sigma_i\ot\sigma_i/4$ that becomes the Ising-type interaction
$
J\sigma_z\otimes\sigma_z/4
\label{eq:coupling}
$
in the weak coupling limit \cite{Levitt}. 
Hereafter we assume $J>0$ so that $\Theta=JT\ge0$, where $T\ge0$ is
the execution time required to implement $S(\Theta)$.

%It recently turns out that any two-qubit interaction can be reduced to the 
%Ising-type interaction by applying appropriate one-qubit pulse sequences \cite{Bremner04, Hill07}.

Any multi-qubit unitary operations can be written as quantum circuits 
composed of
one-qubit unitary gates and the controlled-NOT (CNOT) gates \cite{Barenco95}.
For one-qubit gates in liquid-state NMR quantum computation, 
the most important systematic errors are
pulse length errors and off-resonance errors \cite{Jones09}. These
errors are suppressed with composite one-qubit gates 
\cite{Alway07,Ichikawa11, Bando12}. The remaining task is to construct
robust two-qubit gates.

For two-qubit gates, an error in the parameter $\Theta$ must be considered. 
This error may be caused by (1) inaccuracy of the numerical value of the 
coupling $J$ or (2) influence of other qubits.
It might be caused by an error in gate execution time $T$, too.
Many composite gates have been proposed to suppress this error
\cite{Jones03, phtr, Hill07, Testolin07, Tomita10}.
The composite gates proposed in \cite{Jones03} compensates for the error 
terms for $S(\Theta)$ up to the second order, whereas those proposed in \cite{Hill07} are applicable
to more general interactions. 
However, these merits are realized at the 
cost of longer execution time, which leads to less tolerance to stochastic noise:
Shorter composite pulse is desirable.

All these two-qubit composite gates are designed following the idea 
that any one-qubit quantum circuit robust against the pulse length error
is mapped to a two-qubit composite gate robust against the error
in $\Theta$. The mathematical basis of this approach is that 
${\rm SU}(2)$ may be mapped to a proper subgroup of ${\rm SU}(4)$.  
Although this approach was successful, we might have ruled out 
robust two-qubit gates that have no corresponding one-qubit gates. 

\begin{figure}[b]
   \centering
   \includegraphics[width=3.2in]{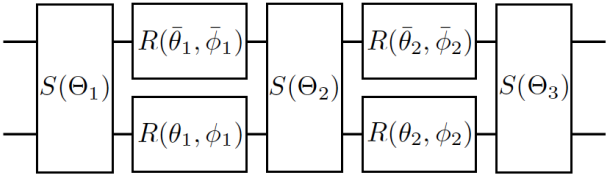} 
   \caption{Schematic diagram of the quantum circuit (\ref{U}) for $N=3$.
   %Quantum gates 
   $R(\theta_i, \phi_i)$ and $S(\Theta_i)$ are
   defined by Eq.~(\ref{R}).}
   \label{QC}
 \end{figure}

In this paper, we start from the most generic
quantum circuit for a two-qubit gate 
in NMR quantum computation depicted in Fig.~\ref{QC}.
Then, we construct minimal robust composite
two-qubit gates. We claim our gates are ``minimal'' in that
the obtained quantum
circuits are constructed with a minimal number of two-qubit gates. 
Moreover, it will be shown below that all our minimal robust composite
two-qubit gates have the corresponding one-qubit composite gates.

%%%%%%%%%%%%%%%%%%%%%%%%%%%%%%
%%%%%%%%%%%%%%%%%%%%%%%%%%%%%%

This paper is organized as follows. In Sec.~\ref{RCond}, we introduce
a circuit family that we are going to analyze.  A robustness condition is
also derived. In Sec.~\ref{RCirc}, we apply it to the circuit family and 
obtain the simplest exact solutions. This solution realizes the robust 
CNOT gates, but not the robust SWAP gates. However, we will find 
a suitable combination of the solutions, which makes the robust gates
equivalent to the SWAP gates up to one-qubit gates. In Sec.~\ref{sec:mapping},
we demonstrate an intriguing property of the solutions. All the composite 
two-qubit gates designed with our solutions have correspondence 
to one-qubit composite gates. By using this, we show that the solutions 
can be seen as extensions of well-known one-qubit composite 
gates. Section~\ref{CDis} is devoted to conclusion and discussions.

%%%%%%%%%%%%%%%%%%%%%%%%%%%%%%
\section{Robustness Conditions}
\label{RCond}
%%%%%%%%%%%%%%%%%%%%%%%%%%%%%%
%\setcounter{equation}{0}

Consider a circuit which implements the following unitary operation, 
\begin{eqnarray}
 U &=& S_N(\bar R_{N-1} \otimes R_{N-1})S_{N-1} \cdots (\bar R_{1} \otimes
  R_{1})  S_1, 
\label{U}
\end{eqnarray}
where $R_i=R(\theta_i, \phi_i)$, $\bar{R}_i=R(\bar{\theta}_i, \bar{\phi}_i)$
and $S_i=S(\Theta_i)$.
For this circuit, we discuss a systematic error in $S(\Theta)$ so that the
real one is given by
\begin{eqnarray}
S'_j = S'(\Theta_j) = S\left((1+\epsilon) \Theta_j \right),%\nonumber
\end{eqnarray}
where $\epsilon\in\R$ quantifies error magnitude. 
We assume here that all the one-qubit gates $R_j$ ($\bar{R}_j$) in Fig.~1 are
implemented instantaneously and error-free, since we can employ robust one-qubit
gates \cite{Alway07, Ichikawa11, Bando12}. On these assumptions, the circuit we actually
implement is no longer $U$, but
\begin{eqnarray}
 U^\prime &=& S_N^\prime(\bar R_{N-1} \otimes R_{N-1})S_{N-1}^\prime \cdots (\bar R_{1} \otimes
  R_{1})  S_1^\prime.%\nonumber
  \label{Uprime}
\end{eqnarray}

In practice, the error strength is unknown but assumed 
to be reasonably small so that the perturbation theory is applicable ($|\epsilon|\ll1$). 
Therefore, we may expand $U^\prime$ as
\be
{U}^\prime=U-i\epsilon \delta U+{\cal O}(\epsilon^2).%\nonumber
\ee
We require the condition
\begin{eqnarray}
\delta U=0
\label{RC}
\end{eqnarray}
for $U'$ to be robust against the error up to the first order in $\epsilon$
and call this condition the robustness condition for $U$.
Note that $U$ and  
$(\bar R_N \otimes R_N) U (\bar R_{0} \otimes  R_{0})$
have the same robustness
since all the one-qubit gates are assumed to be error-free.

%%%%%%%%%%%%%%%%%%%%%%%%%%%%%%
\section{Robust Circuits}
\label{RCirc}
%%%%%%%%%%%%%%%%%%%%%%%%%%%%%%
%\setcounter{equation}{0}
%From now on, 
We will construct minimal circuits that implement nontrivial robust 
two-qubit gates. We are especially interested in a robust $S(\Theta)$ gate
required to implement the CNOT gate. 

%%%%%%%%%%%%%%%%%%%%%%%%%%%%%%
\subsection{No Robust Entanglers with $N$=2}
\label{n2}
%%%%%%%%%%%%%%%%%%%%%%%%%%%%%%

Let us begin our analysis with the case $N=2$, which is the simplest one
in terms of the circuit complexity.
To derive the explicit form of the robustness condition (\ref{RC}), let us 
approximate $S^\prime_j=(1-i\epsilon\Theta_j\sigma_z\otimes\sigma_z/4)S_j$ to the first order in $\epsilon$.
Substituting this to Eq.~(\ref{Uprime}) and using $R_i\sigma_z=\sigma_zR_i^\dag$, we find
\be
\delta U
&=&\frac{1}{4}\[\Theta_2\sigma_z^{\otimes2} S_2(\bar{R}_1\otimes R_1)S_1
+\Theta_1S_2(\bar{R}_1\otimes R_1)\sigma_z^{\otimes2}S_1\]\nonumber\\
&=&\frac{1}{4}\sigma_z^{\otimes2}S_2\(\Theta_2\bar{R}_1\otimes R_1+\Theta_1\bar{R}_1^\dag\otimes R_1^\dag\)S_1.
\ee
Thus, the robustness condition (\ref{RC}) reduces to
\begin{eqnarray}
\Theta_2 (\bar R_1 \otimes R_1)^2 + \Theta_1 \openone^{\otimes2} = 0,%\nonumber
\label{simpleRC2}
\end{eqnarray}
where $\openone$ is the $2\times2$ identity matrix.
The solution is $\bar R_1^2 \otimes R_1^2 = \pm\openone^{\otimes2}$ 
with $\Theta_1/\Theta_2 = \mp1$. Then $U = \pm\bar R_1\otimes R_1$,
which implies that
all the robust circuits belong to $\rm{SU}(2)^{\ot2}$. 
This proves that a robust $S(\Theta)$ for  generic $\Theta$ with $N=2$ is impossible.
We should try $N \geq 3$ to implement a non-trivial two-qubit
gate robust against the error in $\Theta$.

%%%%%%%%%%%%%%%%%%%%%%%%%%%%%%
\subsection{Minimal Robust Entangler with $N$=3} %and its Application to an $N=6$ Robust SWAP Gate}
%%%%%%%%%%%%%%%%%%%%%%%%%%%%%%
The error term $\delta U$ for $N=3$ is written explicitly as
\begin{eqnarray}
\delta U &=& \frac{1}{4}[\Theta_1 \sigma_z^{\otimes2} S_1(\bar R_1 \otimes R_1) S_2 (\bar R_2 \otimes R_2) S_3 \nonumber\\
 &+& \Theta_2S_1(\bar R_1 \otimes R_1) \sigma_z^{\otimes2} S_2 (\bar R_2 \otimes R_2) S_3 \nonumber\\
 &+& \Theta_3S_1(\bar R_1 \otimes R_1)  S_2 (\bar R_2 \otimes R_2) \sigma_z^{\otimes2} S_3].
\end{eqnarray}
Similarly to the $N=2$ case, by using $R_i\sigma_z=\sigma_zR_i^\dag$,
the robustness condition (\ref{RC}) reads
\begin{equation}
\Theta_3 (\bar R_2 \otimes R_2)^2 + \Theta_2 \openone^{\otimes2} 
+ \Theta_1S_2 (\bar R_1^\dagger \otimes R_1^\dagger)^2 S_2^\dagger = 0.
\label{simpleRC}
\end{equation}
We set $\phi_2=\bar{\phi}_2=0$ without loss of generality since we can apply 
$e^{i\bar\xi\sigma_z}\ot e^{i\xi\sigma_z}$ to $U$ and freely adjust 
$\phi_2$ and $\bar{\phi}_2$ without changing the robustness condition.
The robustness condition (\ref{simpleRC}) reduces to
\begin{subequations}
\begin{eqnarray}
\Theta_3 \bar c_2c_2  &=& -\Theta_2-\Theta_1 \bar c_1c_1,
\label{1}\\
e^{-i\phi_1}\Theta_3\bar s_2s_2&=&-e^{i\bar\phi_1}\, 
\Theta_1\bar s_1s_1= -e^{-i\bar\phi_1}\,\Theta_1\bar s_1s_1,
\label{2}\\
e^{-i\phi_1}\Theta_3\bar c_2s_2&=&e^{i\Theta_2/2}\,
\Theta_1\bar c_1s_1 =  e^{-i\Theta_2/2}\Theta_1\,\bar c_1s_1,
\label{3}\\
e^{i\bar\phi_1}\Theta_3\bar s_2c_2 &=& e^{i\Theta_2/2}\Theta_1\,
\bar s_1c_1= e^{-i\Theta_2/2}\Theta_1\,\bar s_1c_1,
\label{4}
\end{eqnarray}
\label{123}
\end{subequations}
\hspace{-3pt}where we introduced $c_i=\cos\theta_i$, $\bar
c_i=\cos\bar\theta_i$, $s_i=\sin\theta_i$, $\bar s_i=\sin \bar\theta_i$
for $i=1,2$. General solutions of Eqs.~(\ref{123}) are  
\begin{eqnarray}
\phi_1=\phi_2=&\bar\phi_1&=\bar\phi_2=0,\nonumber \nonumber \\
\bar\theta_1\equiv\bar\theta_2\equiv 0 \pmod{\pi}, &&%\hspace{2ex}
\Theta_2\equiv0 \pmod{2\pi},\nonumber \nonumber \\
\Theta_1=\frac{\alpha\Theta_2
 \sin\theta_2}{\sin(\gamma\theta_1+\theta_2)},
&& % \hspace{2ex}
\Theta_3=\frac{\beta\Theta_2  \sin\theta_1}{\sin(\gamma\theta_1+\theta_2)}.
\label{gsol}
\end{eqnarray}
For brevity, we introduced symbols
$\alpha, \beta, \gamma\in\{\pm1\}$, which are determined by
$\bar\theta_1, \bar\theta_2$ and $\Theta_2$ (See TABLE \ref{parity}).
Derivation of the above solution is given in the Appendix.
Note that $\Theta_2 \ne 0$ for $U=S(\Theta)$ to be nontrivial.
We list the properties of the solution (\ref{gsol}):
First, the axes of the one-qubit gates are aligned: $\phi_i=\bar\phi_i=0$.
Second, this circuit has the shortest execution time when $\Theta_2=2\pi$, 
since
the execution time $T$ is proportional to $\sum_{i=1}^3 |\Theta_i|
\propto |\Theta_2|$.
We also obtain the other class of general solutions by renaming the qubits as
$1 \leftrightarrow 2$.

%%%%%%%%%%%%%%%%%%%%%%%%%%%%%%
\begin{table}[b]
\caption{Two-value variables $\alpha, \beta, 
\gamma\in\{\pm1\}$ in Eq.~(\ref{gsol}) as functions of $\bar\theta_1,
 \bar\theta_2, \Theta_2$. Note that $\Theta_2\neq0$.}
\centering
\label{parity}
\begin{tabular}{c@{$\hspace{12pt}$}c}
\hline
$\alpha$ & $\bar\theta_1/\pi$\\
\hline
\hline
1&odd\\
$-1$&even\\
\hline
\end{tabular}
\qquad
\begin{tabular}{c@{$\hspace{12pt}$}c}
\hline
$\beta$ & $\bar\theta_2/\pi$\\
\hline
\hline
$\gamma$&odd\\
$-\gamma$&even\\
\hline
\end{tabular}
\qquad
\begin{tabular}{c@{$\hspace{12pt}$}c}
\hline
$\gamma$ & $\Theta_2/2\pi$\\
\hline
\hline
1&even\\
$-1$&odd\\
\hline
\end{tabular}
%%%
\end{table}
%%%%%%%%%%%%%%%%%%%%%%%%%%%%%%

\begin{figure}[b]
   \centering
   \includegraphics[width=3.2in]{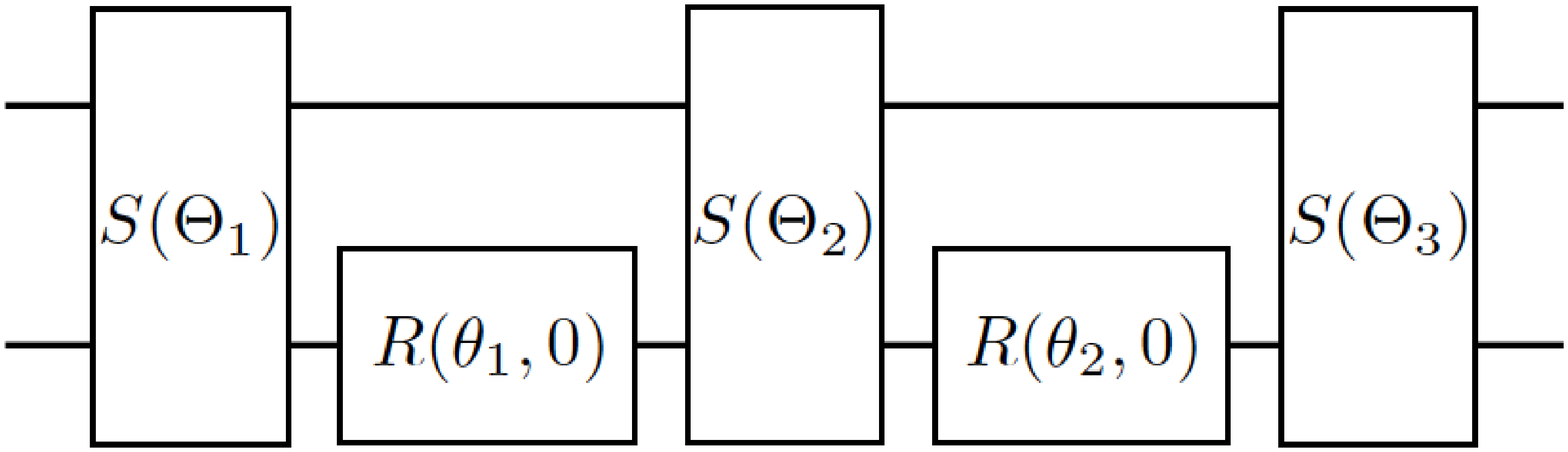} 
   \caption{
Schematic diagram of the minimal robust circuit ($N=3$, $\alpha=-1, \beta=-\gamma$).
%The circuit becomes symmetric and shortest when
  %$\Theta_1 = \Theta_3 =-\pi \sec \theta $, $\Theta_2=2\pi$, $\theta_1=-\theta_2=-\theta$.
   }
   \label{QCSol}
 \end{figure}

The general solutions (\ref{gsol}) are apparently complicated due to various
combinations of $\alpha, \beta$ and $\gamma$. 
Nevertheless, the solution can be simplified, without loss of generality, as 
follows. Let us note that
\be
(\sigma_x \ot \openone )
S\left(\Theta_1\right)|_{\alpha} (\sigma_x \ot \openone )
=
S\left(\Theta_1\right)|_{-\alpha}.%\nonumber
\label{xSSx}
\ee
We can freely flip
$\alpha$ with the one-qubit unitary operations. 
By noting that
$\bar{R}_1=\pm i\sigma_x$ for $\alpha=1$ and $\bar R_1=\pm\openone$ for 
$\alpha=-1$, we can eliminate $\bar{R}_1$ by taking $\alpha= -1$.
Similar observation applies to $\beta$ as well:
$\bar{R}_2=\pm i\sigma_x$ for $\beta=\gamma$
and
$\bar{R}_2=\pm\openone$ for $\beta=-\gamma$.
Thus, we can always take $\alpha=-1$, $\beta=-\gamma$
and the resulting circuit has the minimal number of elementary gates as
shown in Fig.~\ref{QCSol}.
We will work with this circuit from now on.

We present the simplest, and thus the most useful, example that implements 
a composite gate $U=S(\Theta)$ for an arbitrary $\Theta$. 
First, let us take $\Theta_2 = 2\pi$ so that the execution time is
shortest and $\gamma=-1$. 
We then take $\bar\theta_1=\bar\theta_2=0$ to ensure $\alpha=-1$
and $\beta=-\gamma(=1)$.
Further we put $-\theta_1=\theta_2=\theta\in[\pi/2,\pi]$, where $\theta$ is
a parameter chosen in such a way that 
$\Theta_1 = \Theta_3 = -\pi \sec \theta\ge0$
and that $R_1, R_2$ are the shortest pulses.
For this simplest composite pulse $U$, we find the identity
\begin{eqnarray}
[\openone\ot R(\eta,0)]U[\openone\ot R(\eta,0)]^\dag=S(4\zeta),%\nonumber
\label{rus}
\end{eqnarray}
where $\eta$ and $\zeta$ are given by
\be
\tan\eta&=&\tan\theta\sec\(\frac{\pi}{2}\sec\theta\),\nonumber\\
\cos\zeta&=&\cos\theta\sin\(\frac{\pi}{2}\sec\theta\).
\label{eq:zetak}
\ee
Thus, up to one-qubit unitary operations, $U$ implements the operation $S(\pi)$ ($\zeta=\pi/4$).
Since the Cartan decomposition of the CNOT gate contains $S(\pi)$ \cite{NO08},  we observe that
$U$ implements the robust CNOT gate. In case of the shortest pulse, this operation 
is realized by taking
\be
\theta=\pi-\theta^\star,
\qquad
\theta^\star\approx0.674,
%\quad
%n=0,\pm1,\pm2,\ldots, \nonumber
\ee
which has been obtained by solving the second equation of 
Eqs.~(\ref{eq:zetak}) numerically with $\zeta = \pi/4$.

%%%%%%%%%%%%%%%%%%%%%%%%%%%%%%
%%%%%%%%%%%%%%%%%%%%%%%%%%%%%%

This simplest circuit, however, does not implement the SWAP gate.
Note that 
\begin{eqnarray}
S(4\zeta) % &=&{\rm diag}(e^{i \zeta}, e^{-i \zeta}, e^{-i \zeta}, e^{i \zeta})
= \cos \zeta \openone^{\ot2} - i \sin \zeta \sigma_z^{\ot 2},
\end{eqnarray}
while
\begin{eqnarray}
{\rm SWAP}&=&
\left(\begin{array}{cccc}
1 & 0 & 0 & 0 \\0 & 0 & 1 & 0 \\0 & 1 & 0 & 0 \\0 & 0 & 0 &
      1\end{array}\right) %\nonumber 
= 
\frac{\openone^{\ot2}+\sigma_x^{\ot2}+\sigma_y^{\ot2}+\sigma_z^{\ot2}}{2}.\nonumber\\
\label{swapmat}
\end{eqnarray}
Since any one-qubit unitary operations leave the number of 
the summand in the right hand side unchanged,
$S(4\zeta)$ cannot be converted to the SWAP gate, which proves our claim
More precise argument is as follows: First, note that $\{\openone, 
\sigma_x, \sigma_y, \sigma_z\}$ forms an orthogonal basis, in 
the sense of the Hilbert-Schmidt inner product.
By one-qubit unitary transformations, the orthonormal basis undergoes the change $\{\openone,
\sigma_x, \sigma_y, \sigma_z\}^{\ot2}\to V_1\{\openone, \sigma_x, \sigma_y, 
\sigma_z\}W_1\otimes V_2\{\openone, \sigma_x, \sigma_y, \sigma_z\}W_2$,
where $V_i, W_i\in$ SU(2). The new bases $V_i\{\openone, \sigma_x, \sigma_y, 
\sigma_z\}W_i$ are also orthogonal bases. We use these transformed bases to evaluate
the number of summands after the one-qubit unitary operations.
Note that
this can be also shown by using operator Schmidt decomposition 
\cite{Nielsen00, Nielsen03} or SU(2)$^{\ot2}$ invariants \cite{Makhlin02, Zhang03, Zhang04}.
The above observation is proven to be valid also
for the general solutions (\ref{gsol}): The $N=3$ general
solutions generate the robust composite pulses for $S(\Theta)$
only, while there are no such $N=3$ solutions that make the SWAP gate robust.

%%%%%%%%%%%%%%%%%%%%%%%%%%%%%%
\subsection{Minimal Robust SWAP Gates}
\label{swap}
%%%%%%%%%%%%%%%%%%%%%%%%%%%%%%
%\setcounter{equation}{0}

Instead of the SWAP gate, let us consider how to make a composite gate of
the following 2-qubit gate
\be
V
&=&
\left(
\begin{array}{cccc}
1 & 0 & 0 & 0 \\
0 & 0 & -i & 0 \\
0 & -i & 0 & 0 \\
0 & 0 & 0 & 1
\end{array}
\right) %\nonumber 
= 
\frac{\openone^{\ot2}-i\sigma_x^{\ot2}-i\sigma_y^{\ot2}+\sigma_z^{\ot2}}{2}.\nonumber\\
\ee
To investigate the properties of $V$, let us introduce operator Schmidt coefficients
(OSCs) and operator Schmidt number (OSN). The OSCs are coefficients of the basis
operators appearing after the operator Schmidt decomposition is carried out. The OSN stands for
the number of the non-zero OSCs \cite{Nielsen00, Nielsen03}. Note that both the OSCs
and OSN are local unitary (LU) invariants.  This implies that if the OSNs of given two operators
are different from each other, then two operators cannot be locally equivalent.

By using the OSCs, we find that the $V$ gate is LU equivalent to the SWAP gate, since the 
OSCs of $V$ are  $\{1/2,1/2,1/2,1/2\}$, which are identical to those of the SWAP gates
as shown in Eq.~(\ref{swapmat}).
Thus we may identify the $V$ gate with the SWAP gate, up to LU operations.

Now let us evaluate how many elementary $S(\Theta)$ gates we need in order to construct the $V$ 
gate. Clearly it is impossible to construct the $V$ gate (and hence the SWAP gate) with only one elementary $S(\Theta)$ 
gate, as shown in the previous section. We can construct the $V$ gate with two elementary $S(\pi)$ 
gates as
\be
V&=&R\(\tfrac{\pi}{2},\tfrac{\pi}{2}\)^{\otimes2}S(\pi)[R\(\tfrac{\pi}{2},\tfrac{\pi}{2}\)^{\otimes2}]^\dag\nonumber\\
&\times& R\(\tfrac{\pi}{2},0\)^{\otimes2}S(\pi)[R\(\tfrac{\pi}{2},0\)^{\otimes2}]^\dag,
\label{xydec}
\ee
which is a minimal circuit of $V$ within the gate set (\ref{R}).
By replacing two elementary $S(\pi)$ gates in Eq.~(\ref{xydec}) by the $N=3$ composite $S(\pi)$ gates 
(\ref{rus}), we find the $N=2\times3=6$ robust $V$ gate.

Let us check whether there are smaller constructions of a robust $V$ gate based on
the circuit (\ref{xydec}). According to Sec.~\ref{n2}, any $N=2$ robust gate 
implements the trivial gate only. Thus, taking into account the fact that the two factors
the right hand side of Eq.~(\ref{xydec}) implement non-trivial entangling 
gates, it turns out that we must use two $N=3$ solutions (\ref{rus}) for the robust 
implementation of $V$, totaling 6 elementary $S(\pi)$ gates. Furthermore, we find that in the resulting robust $V$ circuit, 
the sub-circuit starting from the third elementary $S(\Theta)$ gate to the fourth cannot be merged
to a single elementary $S(\Theta)$ gate with an appropriate value of $\Theta$, since the OSN of the 
sub-circuit is 4 whereas that of the $S(\Theta)$ gate is 2. This observation shows that 
there exists no reduction to 
an $N=5$ robust circuit. Next, we examine whether there is a reduction to an $N=4$ robust 
circuit. To this end, it is sufficient to consider
whether the sub-circuit starting from the second elementary $S(\Theta)$ to the fourth can be
merged to one elementary $S(\Theta)$ gate. Since the second $S(\Theta)$ gate is $S(2\pi)=-i\sigma_z^{\otimes2}$,
it cannot alter the OSN. This implies that the sub-circuit under consideration has the same OSN 
as the sub-circuit starting from the third to the fourth; thus, for the same reason as the $N=5$ 
case, we have no reduction to an $N=4$ robust circuit.

%%%%%%%%%%%%%%%%%%%%%%%%%%%%%%
\section{Mapping to One-Qubit Composite Pulses}
\label{sec:mapping}
%%%%%%%%%%%%%%%%%%%%%%%%%%%%%%
%\setcounter{equation}{0}

We can map the solutions (\ref{gsol}) to one-qubit composite 
pulses robust against a pulse length error.
Here, the pulse length error is a systematic error, by which the
rotation angle of a one-qubit gate is shifted as
\be
R(\theta, \phi)\rightarrow R((1+\epsilon)\theta, \phi).%\nonumber
\ee
We consider generators
\be
X = \sigma_z\ot\sigma_z, 
\quad
Y = \sigma_z\ot\ve{u}\cdot\ve{\sigma}, 
\quad
Z=\openone\ot\ve{v}\cdot\ve{\sigma}
\label{xyz}
\ee
of SU(4),
where $\ve{u}=(\sin\Omega, -\cos\Omega, 0)$, 
$\ve{v}=(\cos\Omega, \sin\Omega,0)$ 
and $\ve{\sigma}=(\sigma_x, \sigma_y, \sigma_z)$.
Here $\Omega$ is an arbitrary angle.
Since $X, Y$ and $Z$ satisfy the same commutation relations for the generators of $\mathfrak{su}(2)$, 
we recognize a correspondence between the generators
\be
\sigma_x\leftrightarrow X,
\qquad
\sigma_y\leftrightarrow Y,
\qquad
\sigma_z\leftrightarrow Z.
\label{cor}
\ee
%for example.
%for each $\Phi$.
Therefore, given a one-qubit gate $R(\Theta/2, \theta)$, we can associate a rotation 
operation $\mathfrak{R}(\Theta/2, \theta)$ 
in a subgroup of SU(4) by 
\be
R(\Theta/2,\theta)&\leftrightarrow&\exp\[-i(\Theta/2)(\cos\theta X+\sin\theta Y)/2\]\nonumber\\
&=:&\mathfrak{R}(\Theta/2,\theta),%\nonumber
\label{RRmap}
\ee
under the correspondence (\ref{cor}).
Furthermore, the identity
\be
%R(\theta, \phi)
%\hspace{-5ex}
\mathfrak{R}(\Theta/2, \theta)=
%\lefteqn{ 
\[\openone\ot R(\theta,\Omega)\]S(\Theta)\[\openone\ot
					  R(\theta,\Omega)\]^\dag,
					  %}
\label{vj}
\ee
obtained from Eq.~(\ref{xyz}) shows that
a pulse length error in $\Theta/2$ of $R(\Theta/2,\theta)$
is manifestly mapped to a $J$-coupling error in $\Theta$ of $S(\Theta)$ 
under this correspondence.
Note that we have an infinite number of $\mathfrak{su}(2)$
subalgebras parameterized by $\Omega$ in $\bm{u}$ and
$\bm{v}$ and that we obtain the mapping Jones employed in \cite{Jones03} 
as a special case when $\Omega=\pi/2$.

Solutions (\ref{gsol}) can be
mapped locally to one-qubit composite pulses robust 
against a pulse length error by using this mapping. 
Let us introduce a free parameter $\Phi_1$ and define
\be
\Phi_2=\Phi_1-\theta_1,
\qquad
\Phi_3=\Phi_2-\theta_2,%\nonumber
\ee
which clearly satisfy
\be
R(\Phi_2,0)^\dag R(\Phi_1,0)&=&R(\theta_1,0),\nonumber\\
R(\Phi_3,0)^\dag R(\Phi_2,0)&=&R(\theta_2,0).
\label{rdec}
\ee
%Next, 
By using Eqs.~(\ref{vj}) and (\ref{rdec}), we find for 
a circuit $U$ in Fig.~\ref{QCSol} %with $\alpha=\beta=-1$ 
the following identity,
\be
&&[\openone\ot R(\Phi_3,0)]U[\openone\ot R(\Phi_1,0)]^\dag\nonumber\\
&=&[\openone\ot R(\Phi_3,0)]S(\Theta_3)[\openone\ot R(\Phi_3,0)]^\dag\nonumber\\
&\times&[\openone\ot R(\Phi_2,0)]S(\Theta_2)[\openone\ot R(\Phi_2,0)]^\dag\nonumber\\
&\times&[\openone\ot R(\Phi_1,0)]S(\Theta_1)[\openone\ot R(\Phi_1,0)]^\dag\nonumber\\
&=&\mathfrak{R}(\Theta_3/2,\Phi_3)\mathfrak{R}(\Theta_2/2,\Phi_2)
\mathfrak{R}(\Theta_1/2,\Phi_1).
\label{RRR2}
\ee
Since the mapping (\ref{cor}) preserves the $\mathfrak{su}(2)$ algebra 
and the $J$-coupling error is mapped to the pulse length error, 
the one-qubit circuit obtained by replacing
$\mathfrak{R}$ by $R$ in the circuit (\ref{RRR2})
must be robust against the pulse length error. 

Let us consider the simplest case: 
$ \Theta_2 = 2\pi$, $\bar{\theta}_1 =\bar\theta_2=0$ 
and $\theta_1=-\theta_2 =-\theta$. Then, $\Phi_3=\Phi_1$ and 
we obtain a composite {\it rotation} gate in a subspace of SU(4)
\begin{eqnarray} 
\mathfrak{R}\left(\frac{-\pi \sec \theta}{2}, \Phi_1 \right) 
\mathfrak{R}(\pi, \Phi_2) 
\mathfrak{R}\left(\frac{-\pi \sec \theta}{2}, \Phi_1 \right).
\label{eq:map_s}
\end{eqnarray}
A one-qubit gate obtained from (\ref{eq:map_s}) by replacing 
$\mathfrak{R}$ by $R$ is a composite
pulse sequence called SCROFULOUS \cite{CLJ03}. 
It is easy to see that 
Eqs. (21) and (23) in \cite{CLJ03} are satisfied by (\ref{eq:map_s}). 
The {\it rotation} angle $\Theta/2$ as a composite pulse is $\zeta$
defined in Eqs.~(\ref{eq:zetak}),
while $\Phi_1$ determines the {\it rotation} axis. 
Hence, when mapped, our solutions (\ref{gsol}) can be seen as the most
general family of the minimal quantum circuits robust against the pulse
length error and SCROFULOUS is a special case thereof.

The mapping between one- and two-qubit gates makes the fidelity of our 
two-qubit composite pulse and that of the corresponding one-qubit composite 
pulse exactly identical. Readers who are interested in the fidelity plots on the 
two-qubit gates for the simplest cases may refer to Fig. 2 in Ref. \cite{CLJ03}. 

In contrast, all the $N=6$ robust $V$ gates cannot
be mapped to one-qubit robust circuits, because not all the rotation axes
of one-qubit gates are aligned. More precise account is as follows:
First, recall that the mapping to one-qubit gate is defined through $\Omega$ 
in Eq.~(\ref{vj}). Since we have two kinds of one-qubit
gates $R(\pi/2,\pi/2)$ and $R(\pi/2,0)$ in Eqs.~(\ref{xydec}), we cannot
decompose the robust $V$ gate as Eq.~(\ref{RRR2}) 
in such a way that we use the one-qubit gates with a unique $\Omega$.
This implies that the mapping is not well-defined, showing
the nonexistence of the mapping.

As shown in \cite{Jones03}, one-qubit composite gates, e.g., broadband1 
(BB1) and time-symmetric BB1 can be mapped to $N=4$ and $N=5$ compsite
two-qubit gates, respectively. Although these BB1 analogues compensate
for the error terms  up to the second order, they fail to implement
the SWAP gates, since the (time-symmetric) BB1 implements $R(\Theta,\theta)$,
which is, from Eq.~(\ref{RRmap}), mapped to $\mathfrak{R}(\Theta,\theta)$.
This means that they implement the robust $S(\Theta)$ only, which supports
the fact that we have no $N=4,5$ composite SWAP gates.

Final remarks are in order.
We address that our $N=3$ robust two-qubit $S(\Theta)$ gates are geometric 
quantum gates
that utilize the Aharonov-Anandan phase \cite{Aharonov87}. 
This is because all the two-qubit robust gates obtained here
are mapped to one-qubit composite gates robust against a pulse length error,
which are proved to be geometric gates \cite{phtr, Kondo11}.
In general, however, the robust $V$ gates are not 
necessarily geometric, since there is no mapping to one-qubit 
circuits robust against the pulse length error; they happen to be geometric
if we implement $R(\pi/2,\pi/2)$ and $R(\pi/2,0)$ in Eqs.~(\ref{xydec})
by geometric composite gates robust against the pulse length error.

%Our analysis shows that the robustness against the systematic errors
%implies the geometric quantum gates.

%%%%%%%%%%%%%%%%%%%%%%%%%%%%%%
\section{Summary}
\label{CDis}
%%%%%%%%%%%%%%%%%%%%%%%%%%%%%%

In summary, we have found all two-qubit quantum circuits robust against
the error with respect to $\Theta$ in the minimal setup ($N=3$) and have
shown that these were mapped to one-qubit composite pulses
robust against the
pulse length error. Our analysis clarified that the robustness
condition introduces severe restriction on the circuits realized. 
For example, the SWAP gate cannot be robust in this minimal setting.

Although constructed on the Ising-type interaction, our pulse sequence 
may compensate for errors in generic two-qubit 
interaction, if it is nested with the \lq\lq term isolation sequence\rq\rq~in the similar 
way to \cite{Hill07}. The sequence proposed in \cite{Hill07}
is composed by nesting two two-qubit pulse sequences. One is the term isolation
sequence, which reduces the generic two-qubit interaction to the Ising-type interaction \cite{Bremner04}. 
The other is the pulse sequence proposed in \cite{Jones03}. By replacing
$S(\Theta)$ in the latter sequence with the former, one can compensate for
the error with respect to the generic coupling or the gate execution time.
This nesting is applicable to our sequences and hence the resulting sequence
may compensate for the interaction errors in various physical systems.

Furthermore, we have shown that the $N=6$ composite gate LU 
equivalent to the SWAP gate is composed of two minimal
composite gates. The resulting composite gate is minimal and has 
no counterparts in the one-qubit composite gates; besides, they are 
not geometric in general. These contrasts between the CNOT and 
the SWAP gates are of interest, and should be investigated further.

%%%%%%%%%%%%%%%%%%%%%%%%%%%%%%
\begin{acknowledgments}
We are grateful to an anonymous referee, who suggested that there exist
an $N=6$ robust SWAP gate numerically.
This work is supported by \lq Open Research Center\rq~Project for
 Private Universities; matching fund subsidy, MEXT, Japan.
 YK and MN would like to thank partial supports of Grants-in-Aid for Scientific 
Research from the JSPS (Grant No.~23540470).
MN is also grateful to JSPS for partial
support from Grants-in-Aid for Scientific Research (Grant No.~24320008).
\end{acknowledgments}
%%%%%%%%%%%%%%%%%%%%%%%%%%%%%%

%%%%%%%%%%%%%%%%%%%%%%%%%%%%%%
\appendix
%%%%%%%%%%%%%%%%%%%%%%%%%%%%%%

%%%%%%%%%%%%%%%%%%%%%%%%%%%%%%
%\vspace{2ex}
\section{General Solutions of $N=3$ Robustness Condition}
%%%%%%%%%%%%%%%%%%%%%%%%%%%%%%
\renewcommand{\theequation}{A.\arabic{equation}}
  %redefine the command that creates the equation no.
 \setcounter{equation}{0}  % reset counter 
\label{Ap}
We solve Eqs.~(\ref{123}) to find nontrivial solutions (\ref{gsol}).
%which yields the entanglers.
First, from Eqs.~(\ref{3}) and (\ref{2}), we obtain
\be
\hspace{30pt}\Theta_2=2\pi k,
\qquad
k\in 0,1,2,\ldots,
\label{A1}
\ee
with
\be
\phi_1=\pi m,
\qquad
m\in\Z
\label{phi2}
\ee
and 
\be
\bar\phi_1=\pi l,
\qquad
l\in\Z.
\label{A2}
\ee
%respectively.
By substituting Eqs.~(\ref{A1}) and (\ref{A2}) into Eqs.~(\ref{123}),
we find
\begin{subequations}
\begin{eqnarray}
\Theta_1 \bar c_1c_1  &=& -2\pi k-\Theta_3 \bar c_2c_2,
\label{1-A}\\
(-1)^m \Theta_1\bar s_1s_1&=&-(-1)^l\, \Theta_3\bar s_2s_2,
\label{2-A}\\
(-1)^m \Theta_1\bar c_1s_1&=&(-1)^k\,\Theta_3\bar c_2s_2,
\label{3-A}\\
(-1)^l\Theta_1\bar s_1c_1 &=& (-1)^k\Theta_3\,\bar s_2c_2.
\label{4-A}
\end{eqnarray}
\label{123-A}
\end{subequations}
We derive
\be
\Theta_1\Theta_3s_1s_2\sin\(\bar{\theta}_1+(-1)^{k+l}\bar{\theta}_2\)=0
%\nonumber
\label{ooss}
\ee
by multiplying the right (left) hand side of Eq.~(\ref{2-A}) by the left (right) side of Eq.~(\ref{3-A}). 
The solutions are classified into four cases.
 
%%%%%%%%%%%%%%%%%%%%%%%%%%%%%%
\paragraph*{Case 1.}
%%%%%%%%%%%%%%%%%%%%%%%%%%%%%%
When $\Theta_1\Theta_3=0$, the quantum circuit results in that for
$N=2$ or less.
From the previous argument, we find $U\in{\rm SU}(2)^{\ot2}$; this
case cannot implement $S(\Theta)$.

%%%%%%%%%%%%%%%%%%%%%%%%%%%%%%
\paragraph*{Case 2.}
%%%%%%%%%%%%%%%%%%%%%%%%%%%%%%
When $\Theta_1\Theta_3\neq0$ and $s_1=0$,
we obtain $s_2=0$ by using Eqs.~(\ref{2-A}) and (\ref{3-A}).
By introducing $c_1=(-1)^p$ and $c_2=(-1)^q$ with $p, q\in\Z$,
Eqs.~(\ref{123-A}) reduce to
%\begin{subequations}
\begin{eqnarray}
(-1)^p\Theta_1 \bar c_1  &=& -2\pi k-(-1)^q\Theta_3 \bar c_2,
%\label{1-B}
\nonumber\\
(-1)^{l+p}\Theta_1\bar s_1 &=& (-1)^{k+q}\Theta_3\,\bar s_2.
%\label{4-B}
\label{123-B}
\end{eqnarray}
%\end{subequations}
We solve Eqs~(\ref{123-B}) with respect to $\Theta_1$ and $\Theta_3$
to obtain
\be
\Theta_1 &=& -(-1)^p2\pi k\bar s_2\csc\(\bar{\theta}_2+(-1)^{k+l}\bar{\theta}_1\),\nonumber\\
\Theta_3 &=& -(-1)^{k+l+q}2\pi k\bar s_1\csc\(\bar{\theta}_2+(-1)^{k+l}\bar{\theta}_1\).
\label{solA}
\ee

%%%%%%%%%%%%%%%%%%%%%%%%%%%%%%
\paragraph*{Case 3.}
%%%%%%%%%%%%%%%%%%%%%%%%%%%%%%
When $\Theta_1\Theta_3\neq0$ and $s_2=0$,
we find $s_1=0$ from Eqs.~(\ref{2-A}) and (\ref{3-A}).
Hence this case boils down to Case 2, and the solution
is given by Eqs.~(\ref{solA}).

%%%%%%%%%%%%%%%%%%%%%%%%%%%%%%
\paragraph*{Case 4.}
%%%%%%%%%%%%%%%%%%%%%%%%%%%%%%
When $\Theta_1\Theta_3s_1s_2\neq0$,  we find
\be
\bar{\theta}_1+(-1)^{k+l}\bar{\theta}_2=\pi p,
\qquad
p\in\Z\nonumber
\ee
from Eq.~(\ref{ooss}). By using this in Eqs.~(\ref{123-A}),
we obtain
\begin{subequations}
\begin{eqnarray}
\bar c_2\[(-1)^p\Theta_1 c_1 +\Theta_3 c_2\] &=& -2\pi k,
\label{1-C}
%\nonumber
\\
\bar{s}_2[(-1)^{k+m+p}\Theta_1s_1-\Theta_3s_2]&=&0,
\label{2-C}
%\nonumber
\\
\bar{c}_2[(-1)^{k+m+p}\Theta_1s_1-\Theta_3s_2]&=&0,
\label{3-C}
%\nonumber
\\
\bar s_2[(-1)^{p}\Theta_1c_1 +\Theta_3\,c_2]&=&0,
\label{4-C}
\end{eqnarray}
\label{123-C}
\end{subequations}
since $\bar c_1=(-1)^p\bar c_2$ and $\bar s_1=-(-1)^{p+k+l}\bar s_2$.

We further use the following classification to find the solutions of Eqs.~(\ref{123-C}):

%%%%%%%%%%%%%%%%%%%%%%%%%%%%%%
\paragraph*{4-a.}
%%%%%%%%%%%%%%%%%%%%%%%%%%%%%%
When $\bar s_2=0$, we have $\bar c_2=(-1)^q$ with $q\in \Z$ and
\be
(-1)^{k+m+p}\Theta_1s_1-\Theta_3s_2=0\nonumber
\label{osos}
\ee
from Eq.~(\ref{3-C}). Together with Eq.~(\ref{1-C}) we then find
\be
\Theta_1&=&-(-1)^{p+q}2\pi ks_2\csc\(\theta_2+(-1)^{k+m}\theta_1\),\nonumber\\
\Theta_3&=&-(-1)^{k+m+q}2\pi ks_1\csc\(\theta_2+(-1)^{k+m}\theta_1\).\nonumber\\
\label{sol4a}
\ee

%%%%%%%%%%%%%%%%%%%%%%%%%%%%%%
\paragraph*{4-b.}
%%%%%%%%%%%%%%%%%%%%%%%%%%%%%%
When $\bar s_2\neq0$,  we obtain $k=0$.
%Thus in this case we have Eqs.~(\ref{phi2}), (\ref{A2}), 
%$k=0$ and $\phi_2=\bar\phi_2=0$.
%By using these, 
From Eq.~(\ref{A1}), then we observe that the quantum circuit reduces to that for $N=2$.
Hence $U\in {\rm SU}(2)^{\ot2}$ for the robust circuits.

%%%%%%%%%%%%%%%%%%%%%%%%%%%%%%
%\paragraph{Summary.}
%%%%%%%%%%%%%%%%%%%%%%%%%%%%%

After all we find two classes of nontrivial solutions. 
The first class is summarized in (\ref{gsol}), 
which is obtained in Case 4-a.  
Here, we take $\bar\phi_1=\phi_1=0$ in
Eqs.~(\ref{A2}) and (\ref{phi2}) so that
$l=m=0$ in Eqs.~(\ref{sol4a}),
since other cases reduce to this case by using the identity
$R(\theta, l\pi)=R((-1)^l\theta,0)$. 
We also obtain the other class of general solutions by renaming the qubits as
$1 \leftrightarrow 2$.

%--------------------------%


\begin{thebibliography}{99}
%--------------------------%
%
%
\bibitem{Feynman96}
R. P. Feynman,
{\it Feynman Lectures on Computation},
(Westview Press, Boulder, 1996).
%
\bibitem{NC00}
M. A. Nielsen and I. C. Chuang,
{\it Quantum Information and Quantum Computation},
(Cambridge University Press, Cambridge, 2000).
%
\bibitem{Gaitan07}
{F. Gaitan,
{\it Quantum Error Correction and Fault Tolerant Quantum Computing},
(Taylor and Francis, Boca Raton, 2008).}
%
\bibitem{NO08}
{M. Nakahara and T. Ohmi
{\it Quantum Computing: From Linear Algebra to Physical Realizations},
(Taylor and Francis, Boca Raton, 2008).}
%
\bibitem{Jones11}
J. A. Jones,
{Phil. Trans. R. Soc. A} {\bf 361}, 1429 (2003).
%
\bibitem{Jones09}
J. A. Jones,
{J. Ind. Inst. Sci.} {\bf 89}, 303 (2009).
%
\bibitem{CodyJones12}
N. Cody Jones, R. van Meter, A. G. Fowler, P. L. McMahon, J. S. Kim, Th. D. Ladd, Y. Yamamoto ,
{Phys. Rev. X} {\bf 2}, 031007 (2012).
%
\bibitem{Levitt}
{M. H. Levitt,
{\it Spin Dynamics},
(John Wiley and Sons, New York, 2008).}
%
\bibitem{Barenco95}
A. Barenco {\it et al.},
\prA{52}{1995}{3457}.
%
\bibitem{Alway07}
W. G. Alway and J. A. Jones,
{J. Magn. Reson.} {\bf 189}, 114 (2007).
%
\bibitem{Ichikawa11}
T. Ichikawa, M. Bando, Y. Kondo and M. Nakahara,
\prA{84}{2011}{062311}.
%
\bibitem{Bando12}
M. Bando, T. Ichikawa, Y. Kondo and M. Nakahara,
J.\ Phys.\ Soc. Jpn.\ {\bf 82}, 014004 (2013).
%
\bibitem{Jones03}
J. A. Jones,
\prA{67}{2003}{012317}.
%
\bibitem{Hill07}
C. D. Hill,
\prl{98}{2007}{180501}.
%
\bibitem{Testolin07}
M. J. Testolin, C. D. Hill, C. J. Wellard and L. C. L. Hollenberg,
\prA{76}{2007}{012302}.
%
\bibitem{Tomita10}
Y. Tomita, J. T. Merrill and K. R. Brown,
\njp{12}{2010}{015002}.
%
\bibitem{phtr} 
T. Ichikawa, M. Bando, Y. Kondo, and M. Nakahara, 
{Phil. Trans. R. Soc. A} {\bf 370}, 4671 (2012).
%
%\bibitem{deri}
%
\bibitem{Nielsen00}
M. A. Nielsen,
quant-ph/0011036.
%
\bibitem{Nielsen03}
M. A. Nielsen {\it et. al.},
\prA{67}{2003}{052301}.
%
\bibitem{Makhlin02}
Y. Makhlin,
{Quant. Info. Processing} {\bf 4}, 243 (2002).
%
\bibitem{Zhang03}
J. Zhang, J. Vala, S. Sastry and K. B. Whaley,
\prA{67}{2003}{042313}.
%
\bibitem{Zhang04}
J. Zhang, J. Vala, S. Sastry and K. B. Whaley,
\prl{93}{2004}{020502}.
%
\bibitem{CLJ03}
H. K. Cummins, G. Llewellyn and J. A. Jones,
\prA{67}{2003}{042308}.
%
\bibitem{Aharonov87}
Y. Aharonov and J. Anandan, \prl{58}{1987}{1593}.
%
\bibitem{Kondo11}
Y.~Kondo and M. Bando,
{J. Phys. Soc. Jpn.} {\bf 80}, 054002 (2011).
%
\bibitem{Bremner04}
M. J. Bremner, J. L. Dodd, M. A. Nielsen and D. Bacon,
\prA{69}{2004}{012313}.
%--------------------%
\end{thebibliography}
\end{document}